# Polaritonic Fourier crystal

Sergey G. Menabde,[1,†] Yongjun Lim,[2,†] Kirill Voronin,[3,†] Jacob T. Heiden,[1] Alexey Y. Nikitin,[3,4,*] Seungwoo Lee,[2,5,6,**] and Min Seok Jang[1,***]

[1]School of Electrical Engineering, Korea Advanced Institute of Science and Technology, Daejeon 34141, Korea
[2]Department of Biomicrosystem Technology, Korea University, Seoul 02841, Korea
[3]Donostia International Physics Center (DIPC), Donostia-San Sebastián 20018, Spain
[4]IKERBASQUE, Basque Foundation for Science, Bilbao 48013, Spain
[5]Department of Integrative Energy Engineering, KU-KIST Graduate School of Converging Science and Technology, and KU Photonics Center, Korea University, Seoul 02841, Korea
[6]Center for Opto-Electronic Materials and Devices, Post-Silicon Semiconductor Institute, Korea Institute of Science and Technology (KIST), Seoul 02792, Korea

[†] These authors contributed equally to this work.
[*]alexey@dipc.org
[**]seungwoo@korea.ac.kr
[***]jang.minseok@kaist.ac.kr

**Abstract**

Polaritonic crystals – periodic structures where the hybrid light–matter waves called polaritons can form Bloch states – promise a deeply subdiffractional nanolight manipulation and enhanced light-matter interaction. In particular, polaritons in van der Waals materials boast extreme field confinement and long lifetimes allowing for the exploitation of wave phenomena at the nanoscale. However, in conventionally patterned nanostructures, polaritons are prone to severe scattering loss at the sharp material edges, making it challenging to create functional polaritonic crystals. Here, we introduce a new concept of a polaritonic Fourier crystal based on a harmonic modulation of the polariton momentum in a pristine polaritonic waveguide with minimal scattering. We employ hexagonal boron nitride (hBN) and near-field imaging to reveal a neat and well-defined band structure of phonon-polaritons in the Fourier crystal, stemming from the dominant excitation of the first-order Bloch mode. Furthermore, we show that the fundamental Bloch mode possesses a polaritonic bandgap even in the relatively lossy naturally abundant hBN. Thus, our work provides a new paradigm for polaritonic crystals essential for enhanced light-matter interaction, dispersion engineering, and nanolight guiding.

**Introduction**

Photonic crystals – periodic structures with periods comparable with the wavelengths of light – that possess a photonic bandgap have been widely applied to manipulate light propagation, including optical fiber-integrated devices for the telecom industry[1,2]. However, the spatial confinement of photons is limited by diffraction, and their momentum is strictly defined by the hosting medium. At the same time, these limitations are nonexistent for polaritons – quasiparticles born of light coupling to the collective oscillations of charges[3]. Particularly, thin layers of van der Waals (vdW) crystals are able to guide polaritons and confine light into a nanometer-scale volume orders of magnitude smaller than the free-space wavelength[4,5]. Furthermore, the evanescent field of such polaritons stretches beyond a thin polaritonic waveguide, exposing them to external manipulations and hybridization[5-9].

For example, the proximity of a two-dimensional (2D) polaritonic waveguide to a highly conductive metal pad leads to a coupling of the polariton to its mirror image and forming a highly confined "image polariton"[10-12]. The momentum of image polaritons increases as the distance to the mirror surface decreases[11,13]. Notably, this phenomenon was leveraged to excite the image (also called 'acoustic') plasmons in graphene to obtain extremely large momentum and access the non-local quantum phenomena at mid-infrared (mid-IR) frequencies[14,15] as well as to realize ultra-sensitive nanoresonators[16]. Similarly, the mid-IR image phonon-polaritons in polar vdW dielectrics deliver a significantly increased field confinement[11,17,18] and provide much stronger light-matter interaction[19].

Despite the clear advantages of polaritons in vdW crystal layers in terms of strong field confinement and momentum tunability, they suffer from a combination of the inherent material-mediated losses[20,21] and the typically large scattering losses at edges and surface defects[11,17,22]. Consequently, the realization of polaritonic crystals with pronounced wave phenomena is a non-trivial task.

Two approaches have been demonstrated so far to create a polaritonic crystal: patterning the polaritonic waveguide[23-27] and patterning the dielectric substrate[28-30], where conventional nanofabrication methods result in sharp material edges. Additionally, a moiré super-lattice in twisted van der Waals crystals is suggested to provide a periodic environment for plasmons in graphene[31], although its unit cell geometry is strictly defined by the atomic structure of graphene.

So far, 2D polaritonic crystals for hyperbolic phonon-polaritons (HPhP) in hexagonal boron nitride (hBN) and alpha-phase molybdenum trioxide (α-$MoO_3$) have been demonstrated to support collective Bloch modes[23-27,30]. However, dielectric vdW slabs of finite thickness support a practically infinite number of polaritonic modes due to their hyperbolic nature[5]. Therefore, a severe scattering of propagating polaritons at the sharp patterned edges leads to an excitation of multiple higher-order modes, so that the band structure of such polaritonic crystals is filled with mode branches[24,30], hindering the band engineering and manifestation of polaritonic bandgaps for individual modes.

Here, we demonstrate a polaritonic Fourier crystal (FC) for mid-IR HPhP in hBN based on the continuously varying metallic Fourier surface which bypasses the aforementioned limitations of conventional "binary" patterning methods[32]. We employ a wafer-scale Fourier surface covered with gold that acts as a substrate for a pristine polaritonic material (Fig. 1a). Since the momentum of an "image mode", $k$, depends on the distance between the mirror and the polaritonic waveguide, polaritons in such an FC experience a harmonic modulation of their wavevector induced by the Fourier mirror, as illustrated in Fig. 1b and c. Most importantly, due to the stronger field confinement in the hBN layer, higher-order modes in such an FC experience lesser modulation depth, while the intermode scattering is minimized due to the adiabatically varying geometry of the structure.

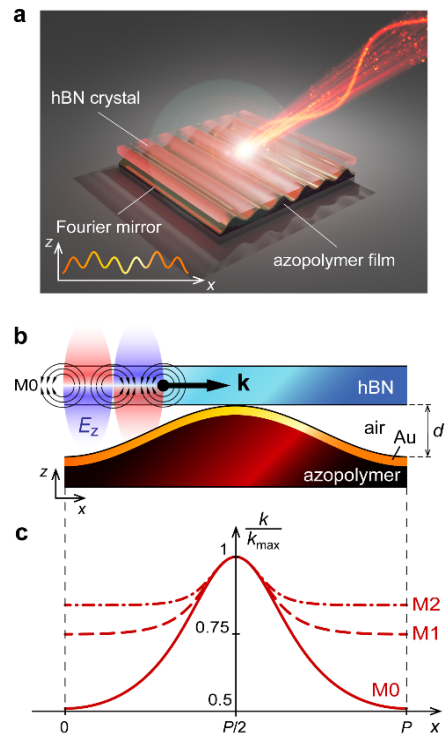

**Figure 1. Polaritonic Fourier crystal. a**, Illustration of a 1D Fourier crystal based on a harmonically corrugated and gilded azopolymer film supporting a pristine polaritonic material. Inset shows the profile of a bi-harmonic Fourier surface used in the drawing. **b**, Schematic of an hBN-based single-harmonic 1D Fourier crystal. When hBN slab is placed on the gilded Fourier surface, the varying distance between hBN and gold induces the spatial modulation of the HPhP momentum in hBN. **c**, Modes of different order experience a significantly different modulation depth across the crystal, thus the manifestation of the polaritonic bandgap is predicted only for the fundamental Bloch mode.

As shown in Fig. 1c for the case of a 100 nm-thick hBN, the momentum of the fundamental (M0) HPhP mode decreases by approximately 50% as the air gap between hBN and gold expands from 0 to 70 nm (also see Methods). At the same time, more confined higher-order modes (M$n$, with $n$ = 1,2,… meaning the number of the field oscillations across the hBN layer) are increasingly less sensitive to the variation of the gap size and thus experience much weaker modulation (less than 25%). Besides, these high-momentum modes are expected to be damped in the absence of coupling mechanisms such as scattering at sharp material edges. Indeed, our near-field experiments and full-wave numerical simulations demonstrate that the polaritonic FC predominantly supports the fundamental Bloch mode originating from the M0 mode in the hBN slab. Furthermore, we numerically demonstrate that the modulation of the momentum of the M0 mode would lead to a manifestation of a wide polaritonic bandgap for this mode even in a relatively lossy naturally abundant hBN crystal[33], used in our experiments. The near-field imaging of the hBN edge suggests a presence of such polaritonic bandgap in the sample.

**One-dimensional polaritonic Fourier crystal for the phonon-polaritons in hBN**

First, a wafer-scale azopolymer film is patterned into a single-harmonic Fourier surface by means of holographic inscription[34] (see Methods). The consequent thermal deposition of a thin (~30 nm) gold layer creates a harmonically corrugated mirror surface. Owing to the phenomenon of image mode, our design provides the first harmonically modulated medium for polaritons in vdW crystals. Furthermore, holographic inscription is a flexible technique, allowing for the fabrication of 1D and 2D Fourier surfaces with tetragonal, hexagonal, quasicrystalline, and multiscale periodicity[34] with an arbitrary Fourier spectrum, opening a vast design space for polaritonic crystals.

Then, a ~300 μm across and 106 nm-thick hBN slab is placed directly on the gilded Fourier surface (Fig. 2a) with the corrugation period $P$ = 485 nm and the height modulation depth $d$ = 70 nm, verified by the atomic force microscope (AFM) scans. The AFM height profile across the hBN edge (Fig. 2b) demonstrates the well-preserved sinusoidal surface of the gilded azopolymer, and that even a 106 nm-thick hBN is rigid enough to remain flat over a span of ≈500 nm. The hBN thickness is selected to provide the polariton wavelengths comparable to $P$ around the middle of the second Reststrahlen band.

Optical properties of the FC are investigated with the scattering-type scanning near-field optical microscope (s-SNOM) which is able to map the near field immediately above the sample with ~10 nm spatial resolution, limited only by the nanotip apex[35]. Examples of the amplitude, $s$, and phase, $\varphi$, of the complex near-field signal $\eta(x,y) = s(x,y)e^{i\varphi(x,y)}$ measured by s-SNOM at different excitation frequencies over the hBN (far from its edges; Fig. 2a) are shown in Fig. 2c. Note that the distributions of both amplitude and phase are drastically frequency-dependent, indicating the different propagation regimes of HPhP in the structure.

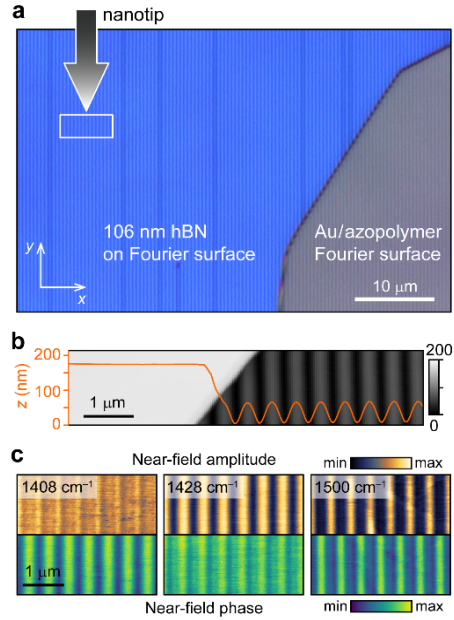

**Figure 2. Fabricated Fourier crystal with hBN. a**, Optical microscope image of a large hBN flake on the gold-covered azopolymer 1D Fourier surface with visible corrugation of 485 nm period. Nanotip schematically shows the near-field scan areas far from the hBN edges. **b**, AFM height profile across the hBN edge, showing the well-preserved sinusoidal profile of the gold surface and flat hBN crystal. **c**, Near-field amplitude (top) and phase (bottom) maps measured by s-SNOM above the sample far from the hBN edges. Both near-field amplitude and phase show strong a dependence on the excitation frequency, revealing different propagation regimes of polaritons in the Fourier crystal.

**Near-field distribution within the Fourier crystal**

In order to correlate our near-field images with the dispersion of HPhP in the FC, we map the near-field amplitude and phase across the frequency range of 1380–1560 cm$^{-1}$ covering most of the second Reststrahlen band of our hBN sample (see Extended Data Fig. 1 and Table 1). Multiple amplitude and phase scans (as in Fig. 2c) are collected with a 4 cm$^{-1}$ frequency shift and averaged along the translation symmetry axis to improve the signal-to-noise ratio. The resulting linear profiles along the $x$-axis are normalized for each scan: amplitude is normalized by its maximum, $s(x)_{\text{norm}} = s(x)/s_{\text{max}}$, and the phase is normalized by its mean value, $\varphi(x)_{\text{norm}} = \varphi(x) - \bar{\varphi}$.

The spectral maps of $s(x, \omega)_{\text{norm}}$ and $\varphi(x, \omega)_{\text{norm}}$ across several unit cells are shown in Fig. 3a. Notably, periodic peaks of the amplitude and phase exhibit a varying width and sometimes drift by $P/2$ along the $x$-axis. This indicates that the recorded near-field signal is dominated by the HPhP in hBN and is not merely due to a profile of the Fourier surface. However, an unambiguous interpretation of the complex near-field signal is a non-trivial task even in uniform structures[36-38]. Furthermore, the near-field is probed far from the hBN edges to guarantee an effectively infinite periodic medium for HPhP.

Thus, the features in the near-field images appearing due to the interference between polaritons in the presence of the material edges are undetectable here, while these features are usually exploited for polariton dispersion analysis in near-field experiments[35,39]. Therefore, we employ full-wave numerical simulations and analytical models to elucidate the origins of the observed near-field patterns and study the HPhP dispersion.

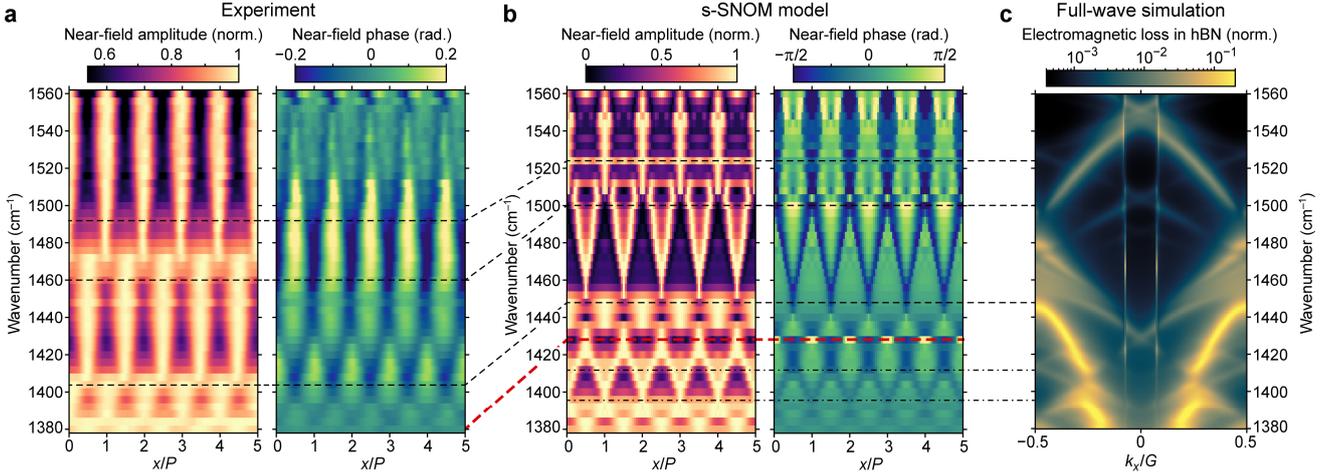

**Figure 3. Spectral map of the near-field distribution. a**, Experimentally measured profiles of the near-field amplitude (left) and phase (right) in the Fourier crystal at different excitation frequencies. Every near-field map is measured across the area of 4×2 μm$^2$, followed by integration to obtain a linear profile, which is then normalized. **b**, Numerically calculated near-field amplitude (left) and phase (right) over the Fourier crystal with the period $P$ = 485 nm and the surface modulation depth 70 nm; $x$ = 0 corresponds to the maximal gap between the hBN and gold surface. Black dashed lines are guides for the eye indicating the corresponding near-field features and showing a blue shift of the calculated data (also highlighted by the red dashed line). **c**, Numerically calculated band diagram of the HPhP modes in the Fourier crystal along the direction of surface corrugation. Dashed and dot-dashed lines indicate a correspondence between the near-field and photonic bands.

We first mimic a s-SNOM experiment by numerically calculating the dipolar moment of a sharp oscillating ellipsoid placed above the hBN in FC at different positions along the $x$-axis (see Methods). Obtained in this manner near-field profiles are normalized in the same way as the experimental data, with origin at $x$ = 0 corresponding to the maximal gap between the hBN and gold as illustrated in Fig. 1b, c. The calculated spectral maps (Fig. 3b) provide a good agreement with the experimental data for both amplitude and phase, indicating the validity of our numerical s-SNOM model. Black dashed lines connecting Figs. 3a and b are guides to the eye showing similar pattern features, and the red dashed line indicates the lower end of the experimentally recorded pattern. A slight blue shift (by ≈ 40 cm$^{-1}$) of the features seen in the numerically-calculated plot with respect to the experimental one can be attributed to the model approximations (e.g. planar HPhP wavefronts in the model instead of the nanotip-launched diverging circular waves[39]).

Then, by employing the full-wave simulations for an infinite FC, the HPhP band structure in the 1D FC is constructed for the Bloch modes with the momentum $k_x$ collinear with the crystal momentum $G = 2\pi/P$ (see Methods). The band structure within the first Brillouin zone in the experimental frequency range is shown in Fig. 3c. Remarkably, despite the infinite number of HPhP quasi-eigenmodes theoretically supported by the hBN slab, the band structure of the FC has only a few well-defined branches. This is expected due to the continuous and adiabatically corrugated gold surface which practically does not cause intermode polariton scattering. In the absence of mechanisms for efficient coupling between the HPhP modes M$n$ of different orders $n$, the fundamental mode M0 is dominant while the inherently more lossy and more energetic higher-order modes are damped. An illustrative comparison between the band diagrams of the FC and a step-like grating is shown in the Extended Data Fig. 2, where HPhP scattering at sharp edges is responsible for generally higher losses and the manifestation of multiple higher-order Bloch modes.

The near-field data (Fig. 3b) and the band diagram (Fig. 3c) are calculated for the same structure, hence a meaningful correspondence can be established between them. For example, the manifesting zig-zag patterns in both amplitude and phase in Fig. 3b appear in the spectral windows between the flat bands – one at the edge (at ≈1450–1500 cm$^{-1}$) and another one in the middle of the Brillouin zone (at ≈1400–1410 cm$^{-1}$). However, a detailed mode analysis is needed to decipher the observed near-field patterns.

**Analysis of individual hyperbolic PhP modes**

In order to study individual modes in the FC, we employ an analytical model which approximates the 3D hBN layer in a periodic medium by a set of 2D sheets of periodically varying conductivity over a flat metallic surface. Each of the bulk modes M$n$ is characterized by its own 2D sheet with the value of the conductivity given by the mode dispersion (see Methods). For simplicity, the conductivity sheets for different modes M$n$ are considered to be independent (no interaction between the modes is considered). For more illustrative results, calculations cover the whole second Reststrahlen band. The band structure in Fig. 4a is an extended data shown in Fig. 3c, and presented here for reference. The individually calculated band diagrams for M0 and M1 modes in the 106 nm-thick hBN are revealed by the poles of the Fresnel reflection coefficient of the periodically modulated 2D conductivity sheets (bright maxima in Fig. 4b). The solid lines indicate the dispersion of the eigenmodes of the uniform conductivity sheet, highlighting the bands bending in FC only for the M0 mode due to its twice larger momentum modulation depth compared to M1 (Fig. 1c).

Despite the simplicity of the analytical model (the fields inside the BN are not considered), it is able to accurately predict the contribution of individual HPhP modes, and shows a very good correspondence with the numerically calculated band structure (Fig. 4a). Firstly, and most importantly, the M0 dispersion exhibits a bandgap opening at the edge of the Brillouin zone, while M1 does not have any bandgaps. Secondly, the crossing of M0 and M1 bands is responsible for the bandgap around 1400 cm$^{-1}$ visible in the full-wave simulations. We note that the analytical model does not account for the scattering between the polaritonic Bloch modes as they form a quasi-normal basis (see Methods). Therefore, a good agreement with the full-wave numerical simulations suggests that the scattering between the Bloch modes due to the surface corrugation is negligible and would not affect the manifestation of the M0 bandgap.

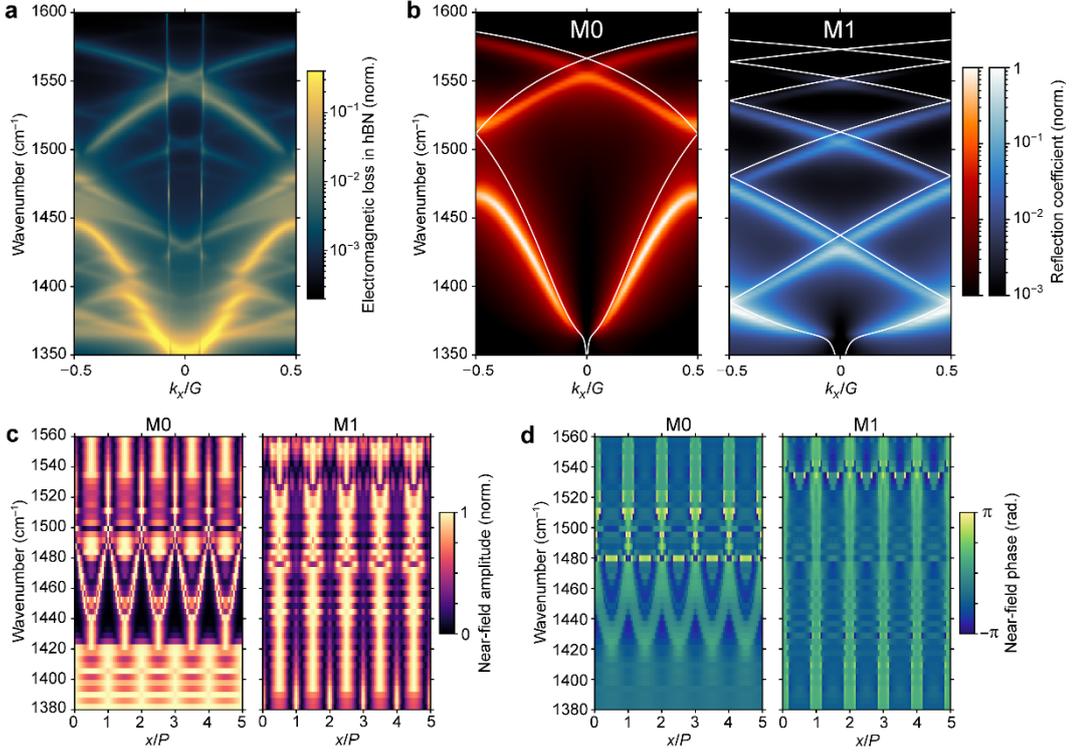

**Figure 4. Analysis of M0 and M1 modes. a**, Numerically calculated band structure of HPhP in the Fourier crystal across the second Reststrahlen band of hBN. Same model is used as in Fig. 3**c**. **b**, Independently calculated band structures of M0 and M1 modes, where the Fourier crystal with hBN is modelled as a 2D sheet of spatially modulated conductivity according to the analytical dispersion of modes in hBN over a conductive surface. Band structures in **a** and **b** are calculated for the direction of surface corrugation. **c**, Numerically calculated near-field amplitude profiles independently generated by M0 and M1 modes. The hBN slab is modelled as a 2D sheet of uniform conductivity which supports only one desired eigenmode, placed over a metallic Fourier surface. **d**, Same as in **c**, but for near-field phase. Both **c** and **d** demonstrate that experimentally and numerically obtained near-field patterns correspond to the M0 mode.

Within our approximation, in order to better understand the role of M0 and M1 modes in the near-field images, we employ the full-wave numerical simulations of the near-field response associated with each mode. This time, unlike Fig. 3b showing calculations for the 106 nm-thick hBN, the hBN is approximated as a 2D sheet of a uniform conductivity (see Methods). Calculated in such a way near-field amplitude and phase distributions corresponding to M0 and M1 modes are shown in Fig. 4c and d, respectively. Spectral patterns originating from the M0 mode are remarkably similar to those observed in the experimental and numerically simulated images (Figs. 3a and 3b, respectively). In contrast, the simulated near-field image originating from the M1 mode (Fig. 4d) lacks any features observed in the experimental data (Fig.3a) and the simulated patterns with all the modes (in Fig. 3b).

Considering the multifaceted analysis in Figs. 3 and 4, we conclude that the experimentally observed near-field response in the FC is essentially dominated by the contribution of the M0 mode. Consequently, this confirms our hypothesis that mostly the fundamental Bloch mode propagates in the FC, in drastic difference from the polaritonic hypercrystals based on "binary" patterned nanostructures where the band diagram is filled with modes without a clearly distinguishable bandgap[30].

However, it is challenging to unambiguously demonstrate the possible presence of the M0 bandgap based on the near-field data in an "infinite" (unrestricted) FC. Meanwhile, a well-defined hBN edge is able to efficiently launch polaritons which interfere with the field under the nanotip and produce near-field fringes with a period approximately equal to the polariton wavelength[17,39-41]. We use this edge-launched mode to study the direction- and frequency-dependent propagation of polaritons in the FC.

**Real-space nanoimaging of the edge-launched polariton**

We investigate the FC area around the hBN edge (Fig. 5a) at frequencies within and outside the expected M0 bandgap, estimated to span from $\approx 1410$ cm$^{-1}$ to $\approx 1470$ cm$^{-1}$ (considering the blue shift of the simulated data in Figs. 3 and 4 by $\approx 40$ cm$^{-1}$). The newly defined $x$-axis is parallel to the hBN edge, and the angle between the $y$-axis and the lattice vector $\vec{G}$ is 32°. Near-field images of the hBN edge (Fig. 5b) are analyzed by taking the fast Fourier transform (FFT; Fig. 5c). The oblique orientation of the edge relative to the corrugation favors distinguishing the weak interference pattern generated by the edge-launched mode in $y$-direction.

The FFT spectra at 1420, 1440, and 1460 cm$^{-1}$ reveal the presence of periodic fringes in $y$-direction (marked by the yellow arrows in Fig. 5c) as expected due to the edge-launched HPhP, illustrated by the yellow wave in Fig. 5a. As the excitation frequency increases, the spatial frequency of the fringes approaches $G$ at 1470 cm$^{-1}$ – the expected upper limit of the M0 bandgap. As the excitation frequency exceeds 1470 cm$^{-1}$, the fringes in $y$-direction practically disappear.

The observed near-field signature of the edge-launched mode can be understood by calculating the band diagram in $y$-direction (Fig. 5d). The predicted M0 bandgap in the direction of $\vec{G}$ (Figs. 4a and b) is shown by the red dashed lines, with the corresponding experimental frequencies (red values and tick marks) considering the 40 cm$^{-1}$ mismatch as discussed earlier. The extracted momentum of the edge-launched mode (yellow circles) agrees well with the M0 dispersion in $y$-direction that has no bandgap.

At the same time, the momentum of the fundamental Bloch mode in $\vec{G}$-direction (red dot-dashed dispersion curves in Fig. 5d) is smaller than the momentum of the M0 mode in *y*-direction. Therefore, when the existence of the fundamental Bloch mode is allowed (at frequencies above 1470 cm$^{-1}$), its excitation by the hBN edge is more efficient than the excitation of the mode in *y*-direction. Thus, experimental results in Fig. 5 indirectly suggest the presence of the M0 bandgap in the direction of $\vec{G}$.

Another notable observation can be made from Fig. 5b: in the presence of the M0 Bloch mode, the near-field amplitude peaks are misaligned with the maxima of the underlying gold surface (dashed lines) identified from the AFM height next to the hBN (Extended Data Fig. 3). In agreement with the discussion in Figs. 3 and 4, this once again indicates that the Bloch mode's near-field dominates the s-SNOM images, outshining even the near-field reflection from the gold surface.

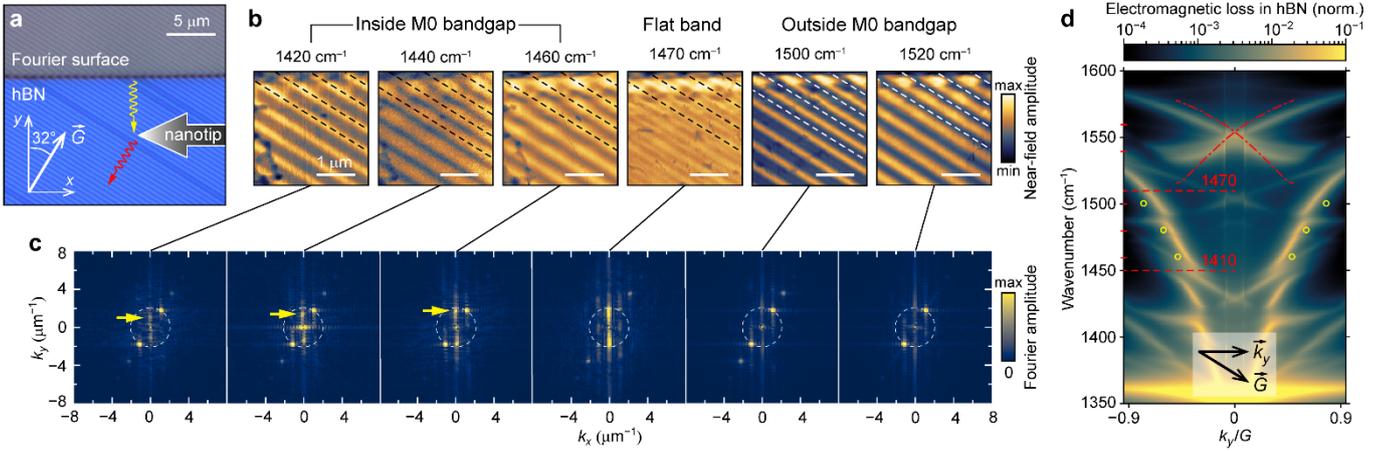

**Figure 5. HPhP probing near the hBN edge. a**, Optical microscope image of the hBN edge in the Fourier crystal, with a coordinate system used for the near-field analysis: the edge is parallel to the *x*-axis and the *y*-axis is at 32° angle to the lattice vector $\vec{G}$. The hBN edge launches the polariton mode in *y*-direction (yellow wave) which produces interference fringes in the near-field images, while the s-SNOM nanotip excites the Bloch modes in the Fourier crystal (red wave). **b**, Near-field amplitude images of the hBN edge shown in **a**, measured at the excitation frequencies around the expected upper limit of the M0 bandgap in the direction of $\vec{G}$. Dashed lines indicate the maxima of the underlying gold surface in contact with hBN. **c**, Corresponding Fourier spectra of the near-field images in **b**. Yellow arrows mark the spectral signature of the near-field interference due to the edge-launched HPhP; the white dashed circle indicates *k* = *G*. **d**, Numerically calculated band structure of the HPhP propagating in *y*-direction (yellow wave in **a**). Yellow circles indicate the experimentally measured momentum of the edge-launched mode: 1.0, 1.25, and 1.65 μm$^{-1}$ at 1420, 1440, and 1460 cm$^{-1}$, respectively. Red tick marks indicate the corresponding experimental frequencies and the red dashed lines show the expected M0 bandgap in the direction of $\vec{G}$ from 1410 to 1470 cm$^{-1}$. Red dot-dashed lines show the analytically calculated M0 dispersion in $\vec{G}$-direction (also shown in Fig. 4b).

**A glance into the bi-harmonic Fourier crystals**

Last but not least, we have designed a bi-harmonic 1D FC for the low-loss isotopically pure hBN[33] where the bandgap opens simultaneously for both M0 and M1 modes (a 'double-bandgap'). These results are shown in Extended Data Fig. 3, and obtained by solving an inverse problem using the spatially modulated 2D conductivity model. It is worth noting that the naturally abundant hBN crystal such as used in our experiments is able to support simultaneous M0 and M1 bandgaps, but only in a system with independent modulation of M0 and M1 modes (Extended Data Fig. 4). However, in real FC, all modes are modulated by a single Fourier surface and cannot be considered independently. Nonetheless, the demonstrated possibility to control the band structure of polaritonic Bloch modes and the possibility of a double-bandgap promise a high application potential for the polaritonic FCs.

**Conclusions**

The concept of a Fourier polaritonic crystal has been proposed and experimentally demonstrated by near-field probing of mid-IR HPhP in hBN within its second Reststrahlen band. First, we show that in spite of the presence of an immense amount of HPhP modes in the hBN layer, the FC possesses a relatively neat and well-defined photonic band structure with a largely dominant fundamental Bloch mode, while no evident signatures of higher-order modes are observed in the near filed. We speculate that this is due to the harmonically corrugated mirror surface that ensures the harmonic modulation of the M0 momentum in hBN by $\approx 50\%$, while the higher-order modes experience significantly lower modulation depth. Furthermore, smooth and continuous surfaces in the structure, including the pristine hBN slab, guarantee a substantial suppression of the intermode scattering for the propagating HPhP Bloch modes. Hence, the high-momentum modes are damped in the absence of the efficient mode coupling mechanism. Second, we numerically demonstrate that the M0 Bloch mode is expected to exhibit a polaritonic bandgap opening even in a relatively lossy naturally abundant hBN crystal. Near-field imaging of the hBN edge hints at the presence of such a bandgap in our sample.

The demonstrated concept of an FC provides a long-sought platform for functional polaritonic crystals where wave phenomena, such as a bandgap opening, can be accessed even for the highly confined hyperbolic polaritons in low-dimensional van der Waals crystals. This promises a plethora of new applications where polaritonic physics can be studied and leveraged for practical use. For example, polaritonic nanocavities based on the FCs potentially can exhibit a large quality factor due to the minimized scattering loss, providing enhanced light-matter interaction, and a low-loss nanolight guiding can be realized akin to the conventional photonic crystals.

**Methods**

**Fourier surface fabrication**

First, we prepared a solution of poly-dispersed red 1 methacrylate (pDR1m) with a molecular weight of 3kDa and a polydispersity index of 1.1-1.2 by dissolving it into 1,1,2-trichloroehtane solvent (Sigma-Aldrich) with 3 wt%. Solution parameters are optimized for the most efficient development of Fourier surfaces[34]. Then, we spin-coated the pDR1m solution onto a 2×2 $cm^2$ silicon wafer, which was cleaned through sequential sonication in acetone, isopropyl alcohol, and deionized water. The spin rate was set to 1000 rpm for 40 seconds, resulting in the azopolymeric thin film with a thickness of 150 nm.

To fabricate a single-harmonic 1D Fourier surface on the azopolymeric thin film, we used an interference pattern generated by the two beams with the left- and right-handed circular polarization. In this process, we used a continuous wave diode laser (Light House Sprout) with a wavelength of 532 nm. The intensity of each beam was set to 150 $mW/cm^2$. Fabrication of the Fourier surface with a period of 485 nm and a modulation height of 70 nm required the inscription time and the beams incident angles of 10 minutes and 33 degrees, respectively.

**Near-field measurements**

Near-field measurements were performed using the neaSNOM from attocube systems AG (formerly Neaspec) coupled with the tunable quantum cascade laser (MIRcat, Daylight Solutions). The Pt-coated AFM nanotips used in s-SNOM (ARROWNCPt, Nano World) had a typical tapping frequency Ω around 270 kHz, and the tapping amplitude was ≈70 nm in a non-contact mode. The background-free interferometric signal[42] demodulated at the third harmonic (3Ω) was used to collect the near-field data.

**Full-wave numerical simulations**

Full-wave numerical simulations of the near-field patterns and the electromagnetic loss density in hBN were performed by the finite-element method in the frequency domain in 2D space using COMSOL Multiphysics software. Both simulation models consider the FC to have a unit cell as illustrated in Fig. 1B. The near-field signal in the mid-infrared range was calculated by applying the electrostatic approach since the grating period and the s-SNOM nanotip curvature radius are deeply subwavelength. The nanotip was approximated by the elongated ellipse with an apex curvature radius of 25 nm. Both nanotip and gold Fourier surfaces were assumed to be a perfect conductor. Thus, we apply a linear potential, $\varphi = -E_0 z$, as a boundary condition for the tip and the metallic grating to emulate the external irradiation. The distance between the tip and the hBN was modulated with an amplitude of 25 nm. Then, the nanotip dipole moment was extracted and demodulated at the third harmonic of the oscillation frequency as a function of the nanotip position and excitation frequency to map the near-field signal. The near-field signals from the M0 and M1 modes were individually obtained by the similar full-wave simulations, but the hBN slab was approximated as a 2D sheet of uniform conductivity which supports only one desired eigenmode with the same momentum as in hBN slab (see below for details).

The band structure of the FC is visualized using the same full-wave numerical simulations in the 2D domain but in periodic configuration by applying the unit cell boundary conditions with Floquet periodicity. Since the gold surface is continuous, it is possible to assume the Otto excitation scheme with a fictitious high-index ($n = 30$) prism underneath the structure which provides the necessary in-plane momentum range in the first Brillouin zone (illustrated in Extended Data Fig. 2). Then, the dispersion of HPhP in hBN and their relative intensity is revealed by the electromagnetic power loss in hBN as a function of the frequency and in-plane momentum of the impinging TM-polarized plane wave in the Otto configuration.

**Analytical model for band structure calculation**

To derive the dispersion relation of HPhP modes in hBN placed at a certain distance above a (flat) metallic surface, we introduce the Cartesian coordinates in which the $xy$-plane is parallel to the hBN slab and the metallic surface. The origin is chosen in a way that the hBN layer is placed between $z = 0$ and $z = t$ planes, and the metal surface is at $z = -d$.

Under the approximation of $k \gg k_0$, the electric field in the system reads:

$$\vec{E}_1 = a_1 \begin{pmatrix} q_x \\ q_y \\ iq \end{pmatrix} e^{i(k_x x + k_y y)} e^{-q k_0 z}; \quad z \geq t$$

$$\vec{E}_2 = a_2 \begin{pmatrix} q_x \\ q_y \\ q_{ez} \end{pmatrix} e^{i(k_x x + k_y y)} e^{i q_{ez} k_0 z} + b_2 \begin{pmatrix} q_x \\ q_y \\ -q_{ez} \end{pmatrix} e^{i(k_x x + k_y y)} e^{-i q_{ez} k_0 z}; \quad t > z \geq 0$$

$$\vec{E}_3 = a_3 \begin{pmatrix} q_x \\ q_y \\ iq \end{pmatrix} e^{i(k_x x + k_y y)} e^{-q k_0 z} + b_3 \begin{pmatrix} q_x \\ q_y \\ -iq \end{pmatrix} e^{i(k_x x + k_y y)} e^{q k_0 z}; \quad 0 > z \geq -d,$$

where $\vec{E}_1$, $\vec{E}_2$, and $\vec{E}_3$ are fields in the air, hBN, and in the space between hBN and metal, respectively, and $q_{ez}^2 = -\frac{\varepsilon_\perp}{\varepsilon_\parallel} q^2$; $q = k/k_0$ is the mode's effective index, $k$ is its momentum, $k_0 = \omega/c$ is that in the free-space, $\omega$ is the excitation frequency, and $c$ is the speed of light. Imposing the boundary conditions at the surfaces of the hBN for the in-plane components, $\vec{E}_{1\parallel}(z = t) = \vec{E}_{2\parallel}(z = t)$, $\vec{E}_{2\parallel}(z = 0) = \vec{E}_{3\parallel}(z = 0)$ and for the out-of-plane components, $\vec{D}_{1\perp}(z = d) = \vec{D}_{2\perp}(z = d)$, $\vec{D}_{2\perp}(z = 0) = \vec{D}_{3\perp}(z = 0)$, and at the metallic surface which is assumed to be perfectly conductive, $\vec{E}_{3\parallel}(z = -d) = 0$, we obtain the following system of equations:

$$a_1 e^{-qk_0 t} = a_2 e^{iq_{ez}k_0 t} + b_2 e^{-iq_{ez}k_0 t}$$
$$ia_1 q e^{-qk_0 d} = a_2 \varepsilon_\| q_{ez} e^{iq_{ez}k_0 d} - b_2 \varepsilon_\| q_{ez} e^{-iq_{ez}k_0 d}$$
$$a_2 + b_2 = a_3 + b_3$$
$$a_2 \varepsilon_\| q_{ez} - b_2 \varepsilon_\| q_{ez} = ia_3 q - ib_3 q$$
$$a_3 e^{qk_0 d} + b_3 e^{-qk_0 d} = 0$$

Or in the matrix form:

$$\begin{pmatrix} e^{-qk_0 t} & -e^{iq_{ez}k_0 t} & -e^{-iq_{ez}k_0 t} & 0 & 0 \\ iqe^{-qk_0 t} & -\varepsilon_\| q_{ez} e^{iq_{ez}k_0 t} & \varepsilon_\| q_{ez} e^{-iq_{ez}k_0 t} & 0 & 0 \\ 0 & 1 & 1 & -1 & -1 \\ 0 & \varepsilon_\| q_{ez} & -\varepsilon_\| q_{ez} & -iq & iq \\ 0 & 0 & 0 & e^{qk_0 d} & e^{-qk_0 d} \end{pmatrix} \begin{pmatrix} a_1 \\ a_2 \\ b_2 \\ a_3 \\ b_3 \end{pmatrix} = 0$$

This linear system has a solution if the determinant of the matrix of this system equals zero. Simplifying the determinant and equating it to zero, we obtain the dispersion relation of the mode:

$$\tan\left(\sqrt{-\frac{\varepsilon_\perp}{\varepsilon_\|}} q k_0 t\right) + \frac{\sqrt{-\varepsilon_\perp \varepsilon_\|}(1 + \tanh q k_0 d)}{1 + \varepsilon_\perp \varepsilon_\| \tanh q k_0 d} = 0$$

This equation cannot be solved explicitly, but if rewritten in the following way it can be easily solved numerically by the iterative approach:

$$q = \sqrt{-\frac{\varepsilon_\|}{\varepsilon_\perp}} \frac{1}{k_0 t} \left[\pi - \operatorname{atan}\left(\varepsilon_\| \sqrt{-\frac{\varepsilon_\perp}{\varepsilon_\|}}\right) - \operatorname{atan}\left(\varepsilon_\| \sqrt{-\frac{\varepsilon_\perp}{\varepsilon_\|}} \tanh q k_0 d\right) + \pi l\right], \quad l = 0, 1, \ldots$$

Note that this equation allows us to find all modes supported by the structure separately. As the high-order modes have a much larger wavevector, the contribution to the observed band structure from them is strongly suppressed. Therefore, we only consider the band structures for the first two modes, $l = 0$ and $l = 1$. On one hand, this approach allows us to separately calculate the bands of different modes to better understand the phenomenon. On the other hand, it has limitations, for example, since the modes are treated independently, the interaction between them is ignored, whereas, in reality, it leads to the manifestation of additional bandgaps as visible in the band structure calculated by the full-wave simulations.

The analytically calculated dispersion of the Bloch modes is revealed by the maxima of the reflection coefficient of the structure, $R$, as a function of frequency and in-plane momentum. To calculate the reflection coefficient as a function of frequency and in-plane momentum, we developed the following procedure. First, we substitute the metallic Fourier surface covered by the hBN crystal with a spatially periodically modulated 2D conductivity sheet. For this, we calculate the wavevector of the mode supported by the hBN slab placed at some distance from the perfect electric conductor and then calculate the 2D conductivity, $\sigma$, supporting the mode with the corresponding wavevector, $\sigma = ic/2\pi q$. Thus, the modulation of the momentum as a function of the distance between hBN and the metallic surface transforms into the modulation of the conductivity of the 2D layer. Such a procedure allows one to avoid considering fields inside the layer, as the latter appears only in the boundary conditions which match the fields above and below the conductivity sheet.

Second, to calculate the reflectance, we use a method developed earlier for the spatially modulated 2D conductivity[43]. Namely, we expand the normalized conductivity, $\alpha = \frac{2\pi}{c}\sigma$, into the Fourier series, $\alpha(x) = \sum_n \bar{\alpha}_n e^{inGx}$, and represent the electric fields in the upper, $\vec{E}_1$, and lower, $\vec{E}_2$, half-spaces by the Fourier–Floquet expansion under the approximation of $k \gg k_0$:

$$\vec{E}_1 = \begin{pmatrix} 1 \\ 0 \\ -i \end{pmatrix} e^{ikx+kz} + \sum_n r_n \begin{pmatrix} 1 \\ 0 \\ i \end{pmatrix} e^{i(k+Gn)x-(k+Gn)z} \; ; \; z \geq 0$$

$$\vec{E}_2 = \sum_n t_n \begin{pmatrix} 1 \\ 0 \\ -i \end{pmatrix} e^{i(k+Gn)x+(k+Gn)z} \; ; \; z < 0$$

Applying the boundary condition to the 2D layer we obtain the following system of equations

$$e^{ikx} + \sum_n (r_n - t_n) e^{i(k+Gn)x} = 0$$

$$e^{ikx} - \sum_n (r_n + t_n) e^{i(k+Gn)x} = 2i \sum_{n,m} \bar{\alpha}_m \frac{k+G(n+m)}{k_0} t_n e^{i(k+Gn)x+imGx}$$

To solve this system, we consider the finite number of harmonics, $n = -M..M$, resulting in the following truncated linear system of equations:

$$\sum_{n=-M}^{M} \left( i\bar{\alpha}_{n-m} \frac{k+Gn}{k_0} + \delta_{nm} \right) t_m = \delta_{n0}$$

$$r_n = t_m - \delta_{n0}$$

Solving this system, we calculate the reflection coefficient $R = \sum_{n=-M}^{M} |r_n|^2$.

**Acknowledgements**

This research was supported by the National Research Foundation of Korea (NRF) grants funded by the Ministry of Science, ICT and Future Planning (NRF-2022R1A2C2092095) and the Ministry of Education (NRF-2021R1I1A1A01057510). S.L. acknowledges support from NRF grants funded by the Ministry of Education (NRF-2022M3H4A1A02074314 and NRF-RS-2023-00272363), Samsung Research Funding & Incubation Center for Future Technology (grant SRFC-MA2301-02), KU-KIST research program (2V09840-23-P023), and Korea University grant. A.Y.N. acknowledges support from the Spanish Ministry of Science and Innovation (grant PID2020-115221GB-C42) and the Basque Department of Education (grant PIBA-2023-1-0007). K.V. acknowledges support from "la Caixa" Foundation (ID 100010434), fellowship code LCF/BQ/DI21/11860026. This work was also supported by the BK21 FOUR Program through the NRF funded by Ministry of Education.

**Author contributions**

S.G.M. and M.S.J. conceived the research idea. S.G.M. fabricated the samples with hBN, performed SNOM measurements, and wrote the manuscript. Y.L. fabricated the Fourier surfaces. K.V. and S.G.M performed numerical and analytical calculations and analyzed the data with valuable input from A.Y.N. J.T.H. assisted in sample fabrication and measurements. A.Y.N., S.L., and M.S.J. supervised the project. All authors contributed to writing the manuscript through discussions and comments.

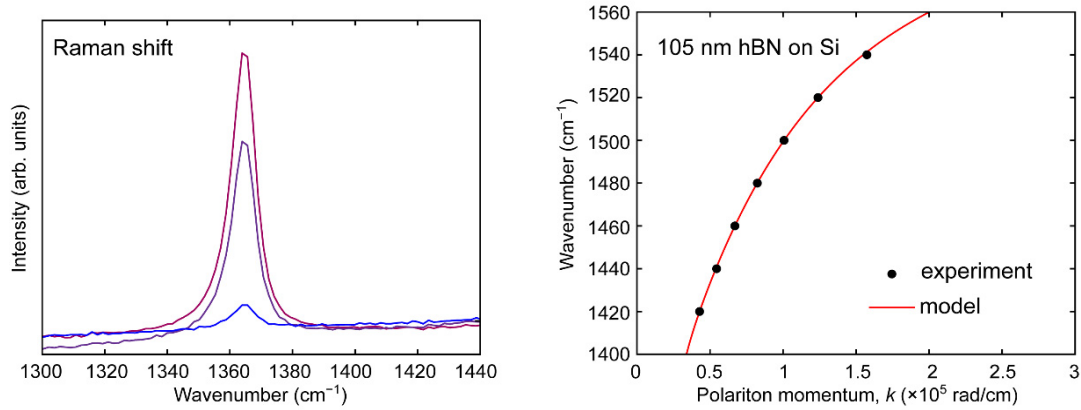

**Extended Data Figure 1.** Left: Raman spectra of three hBN flakes of different thickness, all providing the TO phonon frequency of 1364.3 cm$^{-1}$. Right: HPhP dispersion in 105 nm-tick hBN on Si substrate measured by s-SNOM (data points). Both Raman and near-field data was used to obtain the parameters of the hBN dielectric function provided in Table S1 below. The ~2 nm-thick native SiO$_2$ layer is confirmed to cause insignificant deviation of the HPhP dispersion and is not accounted for during the fitting procedure.

|  | $\omega_{LO}$, cm$^{-1}$ | $\omega_{TO}$, cm$^{-1}$ | $\Gamma$, cm$^{-1}$ | $\varepsilon_\infty$ |
|---|---|---|---|---|
| In-plane | 1619.8 | 1364.3 | 5.3 | 5.22 |
| Out-of-plane | 820.2 | 761.0 | 3 | 2.25 |

**Extended Data Table 1.** Parameters of hBN dielectric function used for modeling. Parameters fitted from the Raman spectroscopy (Fig. S1, left) are shown in blue, and those fitted from near-field dispersion measurements (Fig. S1, tight) are shown in red. The damping parameter is adopted from the literature for the naturally abundant hBN.

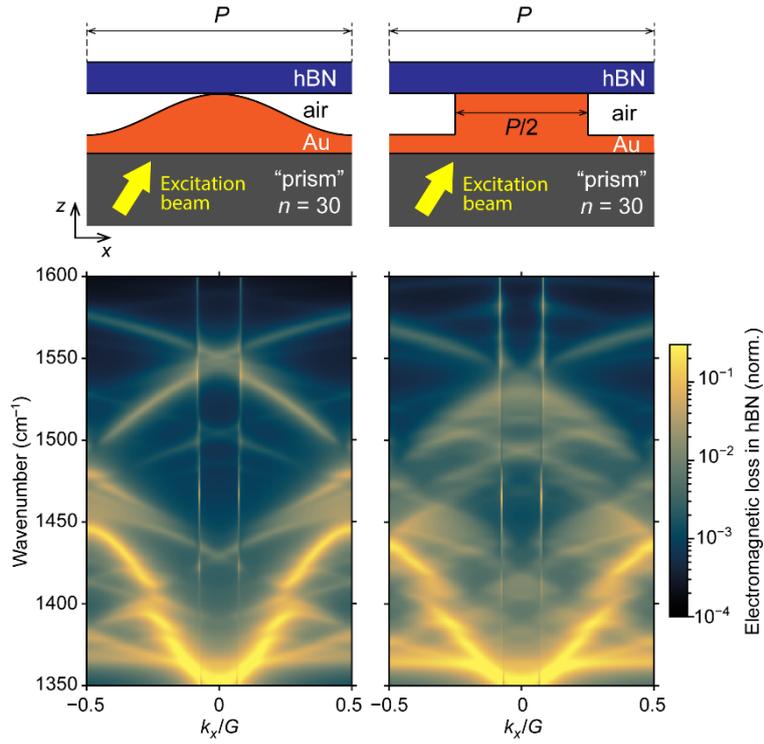

**Extended Data Figure 2.** Band structure in the Fourier crystal (left) and in the grating with sharp edges (right). Both structures have the same maximal air gap size and the period *P*. The gold-hBN contact length in the grating is set to *P*/2. Both band structures are calculated by full-wave numerical simulations in 2D domain with periodic boundary conditions on both sides of the unit cell. Color map shows the electromagnetic loss density in the hBN slab as a function of the excitation frequency and *x*-component of the wavevector of the excitation beam. The excitation scheme mimics the Otto configuration where the TM-polarized plane wave illuminates the structure from below (lower flat interface of the gold film) and is launched by a periodic port immersed in a fictitious high-index prism with $n = 30$. Such a high index is required to sweep through the necessary $k_x$ values with relatively small incidence angles $< 25°$. This provides a relatively uniform excitation efficiency due to the small variation of the reflection coefficient of the flat gold interface.

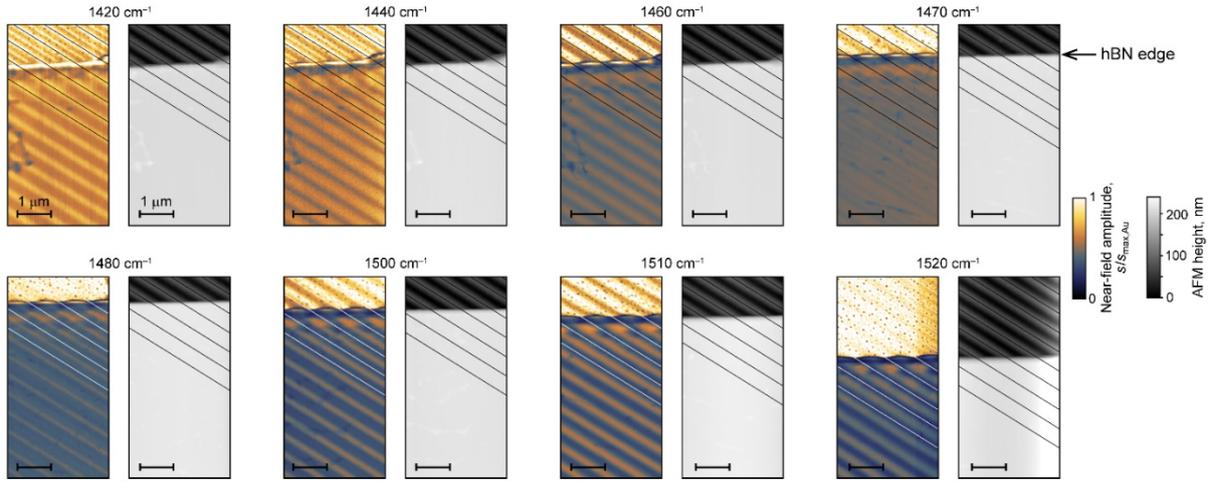

**Extended Data Figure 3.** Near-field amplitude images and corresponding AFM profiles of the hBN edge measured by s-SNOM at different frequencies. Near-field is normalized by the maximum over the gold surface, showing the varying amplitude above the Fourier crystal depending on frequency. Straight lines indicate the maxima of the gold surface where it is in contact with the hBN. Note that the near-field amplitude peaks become misaligned with the gold surface maxima by a half of the corrugation period as soon as the excitation frequency exceeds the upper limit of the M0 bandgap at 1470 cm$^{-1}$.

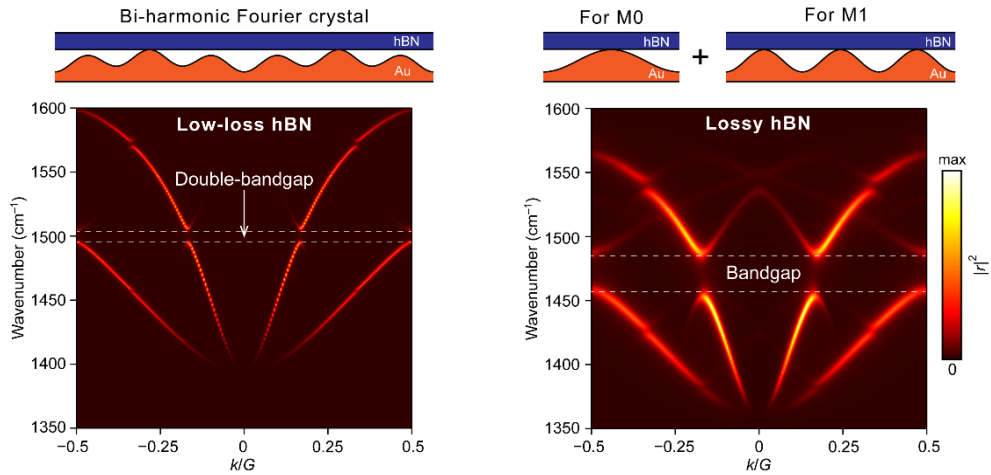

**Extended Data Figure 4.** Left: a double-bandgap opening for M0 and M1 phonon-polariton modes in a bi-harmonic Fourier crystal with 106 nm-thick low-loss isotopically pure hBN. Surface profile is exaggerated. Right: A sum of the two band diagrams calculated independently for M0 and M1 modes in the respective structures with lossy (naturally abundant) hBN. Surface profiles are not in scale.

21